\documentclass[article,twocolumn, prl]{revtex4}
\usepackage{epsfig}
\usepackage{amsmath}
\begin{document}
\title{Nanomechanical displacement detection using
  coherent transport in ordered and disordered graphene nanoribbon resonators} 
\author{A. Isacsson$^*$}
\affiliation{Department of Applied Physics, Chalmers University of
  Technology,
  SE-412 96 G{\"o}teborg Sweden.\\
  $^*$Corresponding author: andreas.isacsson@chalmers.se}
\date{Version: \today}

\begin{abstract}
  Graphene nanoribbons provide an opportunity to integrate
  phase-coherent transport phenomena with nanoelectromechanical systems
  (NEMS). Due to the strain induced by a deflection in a graphene
  nanoribbon resonator, coherent electron transport and mechanical
  deformations couple. As the electrons in graphene have a Fermi wavelength
  $\lambda_F\sim a_0\approx 1.4$~\AA, this coupling can be
  used for sensitive displacement detection in both armchair and
  zigzag graphene nanoribbon NEMS.  Here it is shown that for ordered as well as disordered ribbon systems
  of length $L$, a strain $\epsilon\sim (w/L)^2$ due to a deflection
  $w$ leads to a relative change in conductance $\delta G/G \sim
  (w^2/a_0L)$.
\end{abstract}
 
 \maketitle

 Nanoelectromechanical (NEM) resonators hold
 promise for technological implementations such as tunable
RF-filters 
and ultrasensitive
mass-sensing.
NEMS are also of interest in
connection with fundamental studies of quantum properties of
macroscopic systems.
Regardless of
application area, transduction mechanisms for system control and
readout must be implemented. 

Being only a single atomic layer thick, graphene constitutes the
ultimate material for 2D-NEMS, and graphene NEMS have already been
demonstrated \cite{Mceuen_2007,Bachtold_2008_2, Houston_2008,
  McEuen_2008, Hone_2009, Singh_2010}.  Because electron transport
through mesoscopic graphene devices can be phase
coherent~\cite{deHeer_2006,Morpurgo_2007,Stampfer_2009}, using
graphene in NEMS means that phase coherent transport phenomena can be
directly integrated into NEM resonators. This allows the motion of the
NEMS to couple to the length scale set by the Fermi wave length
$\lambda_F\sim a_0\approx 1.4$~\AA.

\begin{figure}[t]
\epsfig{file=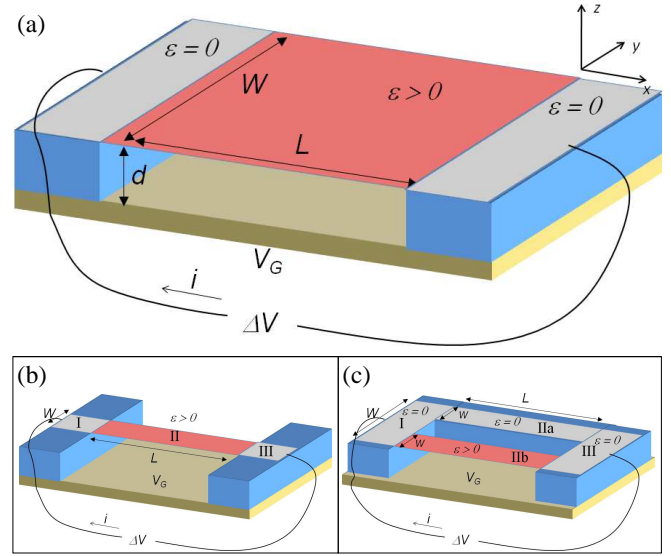, width=\linewidth}\\
\caption{(Color online)
  Suspended graphene NEMS systems. (a) Graphene sheet of length $L$
  and width $W$ suspended in the $xy-$plane above a backgate. (b) Single
  suspended nanoribbon. For armchair nanoribbons, the strain induces a
  transport gap which can be used for displacement detection. (c)
  Graphene nanoribbon interferometer for displacement detection using
  zigzag nanoribbons.\label{fig:systems}}
\end{figure}

So far graphene NEMS have operated in the diffusive transport regime where 
electric~\cite{Hone_2009, Singh_2010} transduction has been based on
charge carrier density modulation. In those experiments the
graphene was suspended above a backgate a distance $d$ as shown in
Fig.~\ref{fig:systems}(a). For a sheet of length $L$ and width $W$
the capacitance to the gate is $C_{\rm G}\approx \epsilon_0
LW/d$~\cite{Bolotin_2008, Andrei_2008}.  Hence, the backgate voltage
$V_{\rm G}$ induces a carrier density $n_0e=\epsilon_0 V_{\rm G}/d$.
This leads to a conductivity of $\sigma=\mu\epsilon_0 V_{\rm
  G}/d$~\cite{Adam_2008}, where $\mu$ is the mobility. Motion
detection then uses the change in carrier density with distance. 
For a deflection $w$ away from the equilibrium distance $d$, the relative change in
conductance is $\delta\sigma/\sigma\sim w/d$~\cite{Blanter_2010}.  Note that only geometric
length scales ($w$ and $d$) enter into this expression.

A deflection $w$ also induces strain $\epsilon\approx (w/L)^2$,
which affects both the dynamical ~\cite{Atalaya_2008, Hone_2009} and
the electronic~\cite{Neto_RMP_2009} properties. For diffusive
transport, strain leads to a linear increase in
resistance~\cite{Hong_2010} with a constant
of proportionality of order unity. 
Hence, the relative change in conductivity due to strain, 
$\delta\sigma_{\rm strain}/\sigma \sim (w/L)^2$, 
is typically negligible compared to that from 
the carrier density modulation (for a more detailed anaysis, see Ref.~\cite{Blanter_2010}). 

The situation is different if one considers coherent transport in
graphene nanoribbons.  The transverse confinement then give rise to
conductance quantization. This prevents conductance changes due to
motion in the backgate electrostatic field leaving strain as the only
coupling between deformation and conductance.  In this paper it is
shown that in such graphene nanoribbon NEM-devices, an operating point
where $\sigma$ changes with displacement as
$\delta\sigma/\sigma=\delta \sigma_{strain}/\sigma \sim (w^2/a_0L)$
can be found.  For armchair nanoribbons [see
fig.~\ref{fig:systems}(b)], this is due to the opening of the
transport gap. For zig-zag nanoribbons, which has no transport gap,
an interferometer type set-up [see Fig.~\ref{fig:systems}(c)] can be
used. This set-up can also be used in the presence of edge-disorder.

Typically, graphene NEMS in equilibrium is not under zero
strain. Not only will this inhibit ripple formation, it will also lead to more
linear mechanical response. This strain can either be due built in strain or due
to biasing to a working point $w_0$. In the latter case, the sensitivity to a
variation $w=w_0+\delta w$ in deflection is naturally linear in  $\delta w$,
i.e. $\delta G/G\sim (w^2/a_oL)\approx (w_0/L)(\delta w/a_0)$. 

Transport through suspended graphene sheets and ribbons and in
graphene with strained regions has been studied previously by several
researchers~\cite{Fogler_2008,Pereira_2009,Leon_2009,Mariani_2009} and
the prospects of using strain in a controlled way to influence
electronic properties is currently an active research
field. Here the focus is on displacement detection in graphene
nanoribbon NEMS.

The electronic properties close to the charge neutrality point are
well described by the nearest neighbor tight binding model.
Suppressing spin indices it is
\begin{equation}\label{eq:TB}
H=-\sum_{n,\delta_i} \left[t_{n\delta_i}a_{n}^\dagger b_{n\delta_i}+{\rm h.c.}\right]+
\sum_{m} [V_{m}^{(a)}n^{(a)}_m+V_{m}^{(b)}n^{(b)}_{m\delta_1}].
\end{equation}
Here $a_n^\dagger$ is the creation operator for an electron at the
point ${\bf R}_n=(n_1{\bf a}_1+n_2{\bf a}_2)$ and $b_{n\delta_i}$ the
destruction operator for electrons at the site ${\bf R}_n+{\bf
  \delta}_i$. The basis is here: ${\bf a}_1={a_0}(3/2,\sqrt{3}/2)$, ${\bf
  a}_2={a_0}(3/2,-\sqrt{3}/2)$, ${\delta}_1={a_0}(1/2,\sqrt{3}/2)$,
${\delta}_2={a_0}(1/2,-\sqrt{3}/2)$, and ${\delta}_3=a_0(-1,0)$. 
For unstrained graphene, $t_ {n\delta}=t_0\approx 2.7$~eV, and
the Fermi velocity is $\hbar v_F=3t_0a_0/2$.

For uniform strain $\epsilon$ along the $x-$direction (armchair edge)
the bond-lengths change from $a_0$ to 
\begin{eqnarray}
|\delta_1|=|\delta_2|=a_0[1+0.25\epsilon(1-3\sigma_p)],\quad
|\delta_3|=a_0[1+\epsilon],\nonumber
\end{eqnarray}
while for uniform strain in the $y$-direction (zig-zag edge),
\begin{eqnarray}
|\delta_1|=|\delta_2|=a_0[1+0.25\epsilon(3-\sigma_p)],\quad
|\delta_3|=a_0[1-\epsilon\sigma_p].\nonumber
\end{eqnarray}
Here $\sigma_p\approx 0.1$ is the Poisson ratio. 

The changed lengths alter the hopping energies as
$t_{n\delta_i}=t_0(1+\Delta_i)$. Typically, $\epsilon\ll 1$ and it
suffices to work to first order in $\epsilon$. As
$\Delta_i\propto \epsilon$, only first order terms in $\Delta_i$
need to be kept.  The spectrum then remains gapless and linear and
can, for uniform strain, be described by the low energy Hamiltonian
$H=H_D+V$ where
\begin{eqnarray}\label{eq:Dirac}
H_D&=&\hbar v_F\left[\begin{array}{cc} \hat\Sigma\cdot(-i\nabla+{\bf k}_0) & 0  \\ 0 & \hat\Sigma^*\cdot(-i\nabla-{\bf k}_0) \end{array}\right]\nonumber.
\end{eqnarray}
Here $\hat{\Sigma}$ is a modified $\hat{\sigma}$-matrix defined as
\begin{eqnarray}
\hat{\Sigma}\equiv v_F^{-1}\left[
\begin{array}{cc} 
0 & {\bf v} \\
{\bf v}^* & 0 
\end{array}\right]=v_F^{-1}\left[
\begin{array}{cc} 
0 & {v_x\hat{x}+v_y\hat{y}} \\
{v_x^*\hat{x}+v_y^*\hat{y}} & 0 
\end{array}\right]\end{eqnarray}
where
\begin{eqnarray}
\hbar v_x&=&\frac{3t_0a_0}{2}\left[1+\frac{\Delta_1+\Delta_2+4\Delta_3}{6}-i\frac{\sqrt{3}}{3}\frac{\Delta_1-\Delta_2}{2}\right]\nonumber\\
\hbar v_y&=&\frac{3t_0a_0}{2}\left[\frac{\Delta_1-\Delta_2}{2\sqrt{3}}-i\left(1+\frac{\Delta_1+\Delta_2}{2}\right)\right]\nonumber\\
{\bf k}_0&=&\frac{1}{3a_0}\left(\sqrt{3}[\Delta_2-\Delta_1],\,\Delta_1+\Delta_2-2\Delta_3\right).\nonumber
\end{eqnarray}
Hence, the Fermi velocity changes and becomes anisotropic, and the locations of
the Fermi points in wavevector space changes.  

The $\Delta_i$:s depend on the direction the strain is applied in and
must be determined from first principles. In the context of carbon nanotubes, this has been studied
extensively~\cite{Kane_1997,Crespi_2000,Han_2000} mainly using
tightbinding H{\"u}ckel theory or Koster-Slater
calculations~\cite{Kleiner_2001}. More recently, density functional
theory has been applied to strained graphene~\cite{Yang_2008,
  Ribeiro_2009}.  Here, the model of Ribeiro et
al.~\cite{Ribeiro_2009} will be used, where 
$t_{\delta_i}=t_0\exp[-\beta_i(\delta_i/a_0-1)]$.

For strain along the $x-$direction (armchair)
$\beta_1=\beta_2=2.6$, $\beta_3=3.3$~\cite{Ribeiro_2009}, and to lowest order in $\epsilon$, one finds
\begin{eqnarray}
\hat{\Sigma}&=&(1-2.35\epsilon)\sigma_x\hat{x}+(1-0.46\epsilon)\sigma_y\hat{y},\label{eq:sigma}\\
{\bf k}_0&=&\frac{2}{3a_0}2.84\epsilon\hat{y}.\quad\quad\quad\quad\quad\quad\quad\quad\quad(armchair)\label{eq:k0ac}
\end{eqnarray}
Here $\sigma_{x,y}$ are the conventional Pauli spin-1/2 matrices. For
strain along the $y-$direction (zig-zag)
$\beta_1=\beta_2=3.15$ and $\beta_3=4$~\cite{Ribeiro_2009}, which
gives
\begin{eqnarray}
\hat{\Sigma}&=&(1-0.74\epsilon)\sigma_x\hat{x}+(1-2.30\epsilon)\sigma_y\hat{y},\label{eq:sigma}\\
{\bf k}_0&=&-\frac{2}{3a_0}2.70\epsilon\hat{y}.\quad\quad\quad\quad\quad\quad\quad\quad(zig-zag)\label{eq:k0zz}
\end{eqnarray}

Consider now an armchair graphene nanoribbon of
length $L$, uniform width $W$, suspended above a backgate as in
Fig.~\ref{fig:systems}(b). The supported parts (regions
I and III) are assumed to be unstrained, while the
suspended part (region II) is under finite strain $\epsilon>0$.
The conductance in the linear response regime is related to the transmission function ${\cal T}(E)$ 
as $G=2(e/h^2){\cal T}(E)$, where the prefactor 2 accounts for spin.

Confinement in the $y$-direction leads to quantization of
transverse wavevector components
$q_{n}=2\pi/3\sqrt{3}a_0+n\pi/W+k_{0}$ for integer $n$. Here
$k_0$ is given by Eq.~(\ref{eq:k0ac}).  If
the interfaces between strained and unstrained regions are along the
$y$-direction transverse mode number $n$ will be conserved. Hence,
${\cal T}=\sum_n{\cal T}_n$ and the problem reduces to solving the 1D
Dirac equation
\begin{equation}
\left[-i\sqrt{\Sigma_x(x)}\partial_x\sqrt{\Sigma_x(x)}+\Sigma_y q_{n}+v(x)\right]\psi_{n}(x)={\cal E}_{n}\psi_{n}(x).
\label{eq:1D_Dirac}
\end{equation} 
Here $v(x)=V(x)/\hbar v_F$ is the effective potential in the ribbon
and ${\cal E}_n=E_n/(\hbar v_F)$.  

To calculate ${\cal T}$ Eq.~(\ref{eq:1D_Dirac}) should be solved in the regions I, II and III [see Fig.~\ref{fig:systems}(b)] and 
the solutions matched at the interfaces (see also Ref.~\onlinecite{Peres2010_RMP}).
In a region of constant $v$ and $\epsilon$, the solution with energy ${\cal E}$ in band $n$ is 
\begin{eqnarray}
\psi_{n}(x)=A_ne^{ikx}\left(\begin{array}{c} 1\\ e^{i\theta_n(k)}\end{array}\right)+B_ne^{-ikx}\left(\begin{array}{c} 1\\ -e^{-i\theta_n(k)}\end{array}\right).
\label{eq:scatt_state}
\end{eqnarray}
where $\exp[i\theta_n(k)]=\frac{k+iq_{n}}{{\cal E}-v}$ and $k=+\sqrt{({\cal E}-v)^2-q_{n}^2}$.

\begin{figure}[t]
\epsfig{file=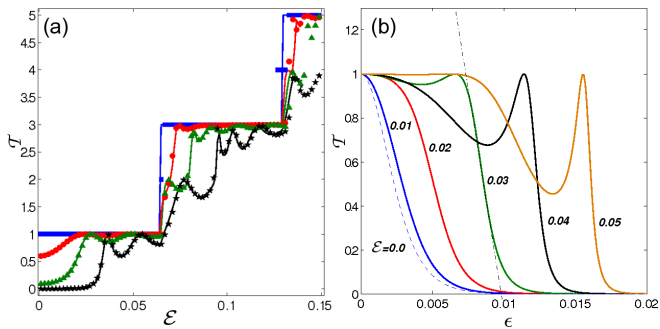, width=\linewidth}
\caption{(Color online)
  (a) Transmission probability ${\cal T}$ through the metallic armchair ribbon
  ($W=17$~nm, $L=28$ nm) [see Fig.~\ref{fig:systems}(b)] as function
  of $\cal E$ and $\epsilon$.  The solid lines follow from the long wavelength approximation
  [Eq.~(\ref{eq:transm})] for the strains $\epsilon=0.0\%$
  (blue\,squares), $0.2\%$ (red circles), $0.5\%$ (green triangles),
  $1.0\%$ (black\,stars). The discrete symbols were obtained using
  numerical tightbinding calculations. (b) Transmission probability ${\cal T}$ as
  function of strain $\epsilon$ for energies ${\cal
    E}=0.0,\,0.01,\,0.02,\,0.03,\,0.04$ and $0.05$. The slope of the
  dash-dotted line determines the sensitivity of the working point
  around $\epsilon=1.0\%$.
\label{fig:comparison1}}
\end{figure}

The matching of wavefunctions between regions is determined by current
conservation~\cite{Silin_1996}.  The current-operator corresponding to
the Hamiltonian in Eq.~(\ref{eq:1D_Dirac}) is
$\hat{J}_x=2v_F\Sigma_x\sigma_x$. Consequently, at the interface
between regions I and II
$\sqrt{\Sigma_x^I}\psi_n^{I}=\sqrt{\Sigma_x^{II}}\psi_n^{II}$.
However, the factors $\sqrt{\Sigma_x}$ cancel in the final expression
for ${\cal T}_n$
\begin{equation}\label{eq:transm}
{\cal T}_n(E)=\left(1+\sin^2\phi_n\left[\frac{\sin\theta_{nI}-\sin\theta_{nII}}{\cos\theta_{nI}\cos\theta_{nII}}\right]^2\right)^{-1}.
\end{equation}
Here $\phi_n\equiv k_{II}L$ while $\theta_{n(I,II)}$ are the propagation angles for electrons in regions I and II. 

In Fig.~\ref{fig:comparison1} (a), ${\cal T}$ is shown for strains
$\epsilon=0.0-1.0\%$ as the solid lines.  The discrete symbols, were
obtained numerically using the tightbinding Hamiltonian in
Eq.~(\ref{eq:TB}) and the relation ${\cal T}={\rm Tr}[\Gamma_L
  G_C^r\Gamma_RG_c^a]$. Here $G_C^{r,(a)}$ are the retarded (advanced)
Green's functions for the ribbon and $\Gamma_{L,R}$ are self energies
accounting for semi-infinite graphene leads.  Also, in the numerical
calculation, no linearization in strain has been made.  As can be
seen, for the lowest plateau, a transport gap opens up with increasing
strain.

From Eq.~(\ref{eq:transm}), the sensitivity of the conductance
$G\propto {\cal T}$ to ribbon displacements $w$ can be obtained.
For the lowest transverse mode $q=0$ and for $v_{I,II}=0$ one finds
\begin{equation}
T_0({\cal E})=\frac{{\cal E}^2-k_0^2}{{\cal E}^2-{k_0^2\cos^2 L\sqrt{{\cal E}^2-k_0^2}}}.
\label{eq:T0}
\end{equation}
 
This dependence of $\cal T$ on $\epsilon$ is shown
in Fig.~\ref{fig:comparison1}(b). Different curves correspond to
different back-gate bias points, i.e. different values of ${\cal
  E}=0.0, 0.01,...$. From this figure it is
clear how for a given strain, one may chose a working point ${\cal
  E}_0$ (by gating the structure) such that the slope of the
${\cal T}(\epsilon)$-curve is maximal. This maximal slope, $|\partial {\cal
  T}/\partial\epsilon|_{\rm max}$ then defines the sensitivity.

The smallest sensitivity obtains for the working point at ${\cal E}=0$
[dashed line in Fig.~\ref{fig:comparison1}(b)]. 
A lower bound for
the sensitivity can be found by setting ${\cal E}=0$ in
Eq.~(\ref{eq:T0}) and solving for the maximum magnitude of the slope.
This gives $\left|{\partial{\cal T}}/{\partial \epsilon}\right|_{\rm
  max}\approx \frac{8}{3\sqrt{3}}({L}/{a_0})$. Hence, for a 
deflection of magnitude $w$, the relative change in conductance is 
$\delta G/G={\delta{\cal T}}/{\cal T}\sim {w^2}/(La_0)$.

\begin{figure}[t]
  \epsfig{file=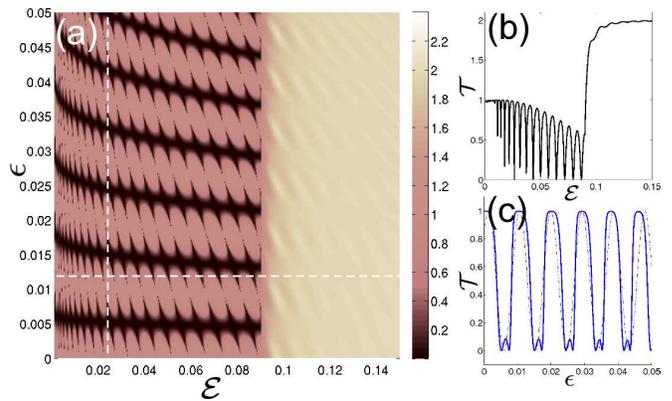,width=\linewidth}
\caption{(Color online)
  (a) Transmission ${\cal T}$ as function of energy $\cal E$ and strain
  $\epsilon$ through the interferometer shown in
  Fig.~\ref{fig:systems}(c).  Broad dark bands correspond to
  destructive interference, whereas
  broad bright bands to unit transmission. The figure was generated
  using the tightbinding Hamiltonian and recursive lattice Green's
  function on a system of width $W=2 \times 7.5$~nm, $L=24$~nm, and a 'gap' between the
  ribbons of $0.5$ nm.  (b) ${\cal T}$ as function of $\cal E$ taken
  along the horizontal dashed line in panel a.  (c) ${\cal T}$ as
  function of $\epsilon$ taken along the vertical dashed line in panel
  (a). The thick solid line is the result of numerical calculation
  whereas the thin dashed line corresponds to the expression
  $\sin\phi=\sin [1.8\epsilon(L/a_0)]$.
  \label{fig:interf}}
\end{figure}

This result is valid for a metallic armchair ribbon where all edges
are perfect and impurities absent.  For transport restricted to the
lowest transverse subband long range impurity scatterers will not
affect the transport~\cite{Wakabayashi_2009}.  However, short range
potentials will have effect.  For armchair graphene nanoribbons both
theory~\cite{Blanter_2009}, and subsequent experiments~\cite{Kim_2009}
suggest that at low temperature, edge disorder induce localization. In
this case, transport at low energies is goverend by variable range
hopping and strongly supressed. Hence, schemes relying on a single armchair ribbon
require nearly perfect edges.

Zig-zag nanoribbons are less sensitive to edge disorder. However,
applying strain will not lead to a transport gap.  Instead, to obtain
a sensitivity of $w^2/(La_0)$ an interferometer with the suspended
ribbon making up one of the arms [see Fig.~\ref{fig:systems}(c)] can
be used. In graphene ring-geometries Aharanov-Bohm oscillations have
been observed at low temperatures~\cite{Russo_2008}.  Here, no
external B-field is required. Instead, the effective gauge field due
to the strain in the suspended arm is exploited. 

\begin{figure}[t]
  \epsfig{file=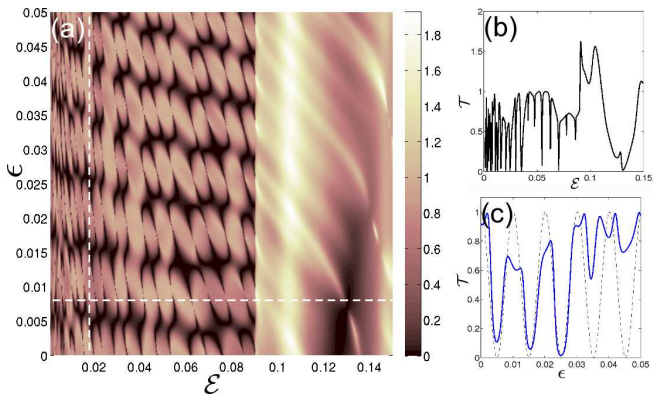, width=\linewidth}
\caption{(Color online)
  (a) Transmission probability ${\cal T}$ as function of energy $\cal E$ and strain
  $\epsilon$ through the same structure as in 
  Fig.~\ref{fig:interf}, but with added edge disorder (probability to remove edge atoms 30\%). 
  (b) ${\cal T}$ as function of $\cal E$ taken
  along the horizontal dashed line in panel a. (c) ${\cal T}$ as
  function of $\epsilon$ taken along the vertical dashed line in panel
  (a). The thick solid line is the result of numerical calculation
  whereas the thin dashed line corresponds to the expression
  $\sin\phi=\sin [1.8\epsilon(L/a_0)]$. 
\label{fig:interf_dis}}
\end{figure}

The idea is again to use the lowest quantized conductance plateau. For
zig-zag nanoribbons this is formed from current carried by the edge
states.  Hence, consider an edge-state coming from the unstrained
region I, which split into the two arms IIa and IIb [see
Fig.~\ref{fig:systems} (c)].  The state propagating in the strained
arm (IIb) will aquire an extra phase $\phi=k_0L=1.8\epsilon (L/a_0)$
[see Eq.~(\ref{eq:k0zz})], and consequently one expects interference
to modulate the conductance with a factor $\propto \sin(\phi)$.

In Fig.~\ref{fig:interf}(a) the result of numerically calculating
$\cal T$ (using the tight-binding Hamiltonian) from region I to region III as function of energy 
${\cal E}$ and strain $\epsilon$ is shown. Bright regions correspond to
${\cal T}=1$ and dark regions to ${\cal T}=0$. These
broad dark regions arise due to destructive interference.  The fine
structure is the result of backscattering at the interfaces where the
ribbon split (inter-valley scattering). This is also visible in Fig.~\ref{fig:interf} (b) where
the transmission for a specific strain $\epsilon=1.2$\% is shown as
function of $\cal E$.

In Fig.~\ref{fig:interf}(c) $\cal T$ is shown for a fixed $\cal E$ as
function of strain (thick solid line).  The period of the conductance
oscillations agree well with the plotted function $1+\sin\phi$ (thin
black line). Hence, approximating ${\cal T}\approx 1+\sin \phi$ yields, as was the case for the armchair ribbon, a sensitivity to
deflections of $\delta {\cal T}/{\cal T}\sim (w^2/La_0)$.  

The effect of edge disorder on the interference pattern is shown in
Fig.~\ref{fig:interf_dis}(a). Here, disorder has been accounted for by
removing the outermost atoms with probability $p=0.3$ at random along
each of the four zig-zag edges of the interferometer. Note that
although $\cal T$ as function of $\cal E$ [panel (b)] is highly
irregular, $\cal T$ as function of strain [Fig.~\ref{fig:interf_dis}
  (c)] show clear conductance modulations. Furthermore, comparing the
solid line and the dashed line in Fig.~\ref{fig:interf_dis} (c), it is
clear that the the expression $\sin\phi=\sin [1.8\epsilon(L/a_0)]$ is still valid.

In conclusion, by exploiting the possibility
to directly integrate coherent electron transport with graphene nanoribbon NEMS, the
conductance through the structure can be made to depend on the mechanical
deflection $w$ as $\delta G/G\sim(w^2)/(La_0)$. This is due to the strain
induced shift of the Fermi-points (synthetic gauge field). It is this shift which causes the
length scale $a_0\sim \lambda_F\sim 1/k_F$ to enter the expression for $\delta G$.

The author wishes to thank J. Kinaret, M. Jonson and
M. Medvedyeva.  This work has received funding
from the Swedish Foundation for Strategic Research and the European
Community's Seventh Framework program (FP7/2007-2011) under grant
agreement no: 233992.

\end{document}